# Ion Beam Effect on GeSe Chalcogenide Glasses: Non-volatile Memory Array Formation, Structural Changes and Device Performance

M. R. Latif[1], T. L. Nichol[1], I. Csarnovics[2], S. Kökényesi[2], A. Csik[3], D.A. Tenne[4], M. Mitkova[1]

1. Department of Electrical and Computer Engineering, Boise State University, Boise ID – USA

2. Department of Experimental Physics, University of Debrecen, Debrecen – Hungary

3. Institute for Nuclear Research, Hungarian Academy of Sciences, Debrecen- Hungary

4. Department of Physics, Boise State University, Boise ID - USA

**Abstract**

The conductive bridge non-volatile memory technology is an emerging way to replace traditional charge based memory devices for future neural networks and configurable logic applications. An array of the memory devices that fulfills logic operations must be developed for implementing such architectures. A scheme to fabricate these arrays, using ion bombardment through a mask, has been suggested and advanced by us. Performance of the memory devices is studied, based on the formation of vias and damage accumulation due to the interactions of $Ar^+$ ions with $Ge_xSe_{1-x}$ (x=0.2, 0.3 and 0.4) chalcogenide glasses as a function of the ion energy and dose dependence. Blanket films and devices were created to study the structural changes, surface roughness, and device performance. Raman Spectroscopy, Atomic Force Microscopy (AFM), Energy Dispersive X-Ray Spectroscopy (EDS) and electrical measurements expound the $Ar^+$ ions behavior on thin films of $Ge_xSe_{1-x}$ system. Raman studies show that there is a decrease in area ratio between edge-shared to corner-shared structural units, revealing occurrence of structural reorganization within the system as a result of ion/film interaction. AFM results demonstrate a tendency in surface roughness improvement with increased Ge concentration, after ion bombardment. EDS results reveal a compositional change in the vias, with a clear tendency of greater interaction between ions and the Ge atoms, as evidenced by greater compositional changes in the Ge rich films. The device performance is influenced by the combination of these structural and compositional alterations caused by ion interaction. Electrical testing of the devices was achieved through a measurement of current-voltage (I-V) curves, Read/Write voltages, and the two resistive states. The advantage of this method for array formation is that it provides a unique alternative to conventional photolithography, for prototyping redox conductive bridge memristor devices without involving any wet chemistry.

## 1. Introduction

The requirement for a continued scaling over the years has brought the semiconductor industry to a point where device physics poses severe scaling challenges for the future [1, 2]. Hence, it is essential to keep increasing the number of transistors on a chip by investigating new technologies, which will allow further scaling of devices at the rate predicted by Gordon Moore

[3, 4]. An imminent scaling limit of Si-based flash memories accelerates the search for new memory technology [5]. Memory technologies such as Ferroelectric Random Access Memory (FeRAM) and Magneto-Resistive Random Access Memory (MRAM) are in production but they cover only a niche market for special applications as they exhibit technological and inherent problems in the scalability [6].

To overcome the problems of current non-volatile memory (NVM) concepts, a number of alternative memory technologies are being explored. NVMs based on electrically switchable resistance have gained considerable attention over the last decade, most commonly known resistive random access memory (RRAM). RRAM piques more and more interest as a promising candidate for next generation NVM and is believed to be a universal memory as it exhibits high speed access, fast switching, high density, and low power consumption [7-10]. RRAM devices have a simple capacitor like Metal insulator Metal (MIM) structure, as depicted in Figure 1. Their performance can be based on phase change occurring in the material between the two electrodes [], by formation of conductive path by oxygen vacancies in different oxides (e.g., NiO [11, 12], $TiO_2$ [13, 14], $HfO_2$ [15-17]) or by formation of a conductive bridge of highly mobile metal atoms (Ag) within a solid electrolyte (e.g., GeS [18, 19], GeSe[10]). The devices of interest in this study are based on the third type mechanism and are usually called Programmable Metallization Cell (PMC). Since the metallic filament in solid electrolyte is formed by redox reaction at respective electrodes it called Rodex Condutive Bridge Memory (RCBM) in this work. The device switches between high resistance state (HRS) and low resistance state (LRS) due to the formation or dissolution of conductive molecular bridge between the two metal electrodes by the application of either Write or Erase voltage. HRS represents the OFF state while LRS represents the ON state of the device. In addition, the Write and Erase voltages do not need to be applied continuously to maintain the state, thus making RCBM a NVM.

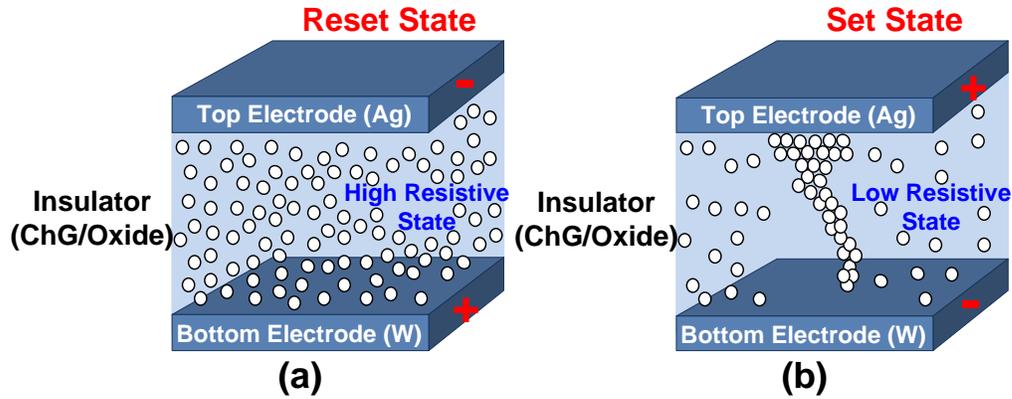

*Figure 1*  RCBM Cell based on Ag doped chalcogenides in (a) High Resistive State (b) Low Resistive State

Traditionally, single bit RRAM cells and arrays are fabricated using photolithography, which involves several steps in completing the device, and also limits the minimum feature size. Success achieved at the single bit level suggest that RRAM is well placed for high performance memory and logic applications. By integrating the RRAM cells into a system, it can fulfill the essential role of memory with high storage density, precision, and access speed [20]. The array structure also provides powerful capability in information processing [21], synapse creation for neuromorphic systems [22], arithmetic computation [23], pattern comparison [24], and reconfigurable field programmable gate array (FPGA) [25].

In this work, a method is proposed for creating $Ge_xSe_{1-x}$ chalcogenide (x = 0.2, 0.3 and 0.4) based RCBM arrays by directly creating the vias in the GeSe active film, by bombarding $Ar^+$ ions through a mask. The openings in the mask thus defined the array size. Formation of array relies on the high resistivity of the GeSe film which expands in the range of hundreds of giga ohms. In this manner the non-sputter regions becomes a neutral isolation between the adjacent cells in the array. The arrival of the energetic ions causes the surface atoms to be removed from the target material and as a consequence it creates vias in the GeSe layer. These high energy particles cause surface roughness, compositional variations, and irreversible structural alterations in the irradiated layer [26]. An investigation on the combination of these effects over $Ge_xSe_{1-x}$ films will allow us to enhance RRAM array performance through process and material optimization. The usability of this method is demonstrated by electrical testing of the RRAM arrays.

## 2. Experimental

Arrays of RCBM devices were created on a $Ge_xSe_{1-x}$/W/SiO$_2$/Si stack (x=0.2, 0.3, 0.4). 200 nm of SiO$_2$ was thermally grown on a Si <100> substrate, followed by 100 nm of sputtered W (Tungsten), and 1 μm of thermally evaporated GeSe chalcogenide thin films. Devices array was created in the chalcogenide layer through $Ar^+$ ion bombardment using an INA-X (SPECS, Berlin) Secondary Neutral Mass Spectrometer (SNMS) by placing a 50μm×50μm nickel mesh over the sample [R. Lovics: Vacuum 86 (2012) 721-723]. In contrary to most ion beam tools where ion energy has a Gaussian distribution, the current is highly uniform within the SNMS machine with an equal distribution of the Ar+ ions over the entire bombarded region. This resulted in uniform depth profiles in the GeSe layer for all the cells in an array. A copper mask was placed on top of the mesh to hold the nickel mesh in place and protect part of the sample from ion bombardment for analysis of ion-induced effects. Surface bombardment was performed with low pressure Electron Cyclotron Wave Resonance (ECWR) $Ar^+$ plasma. The resulting configuration can be seen in Figure 2.

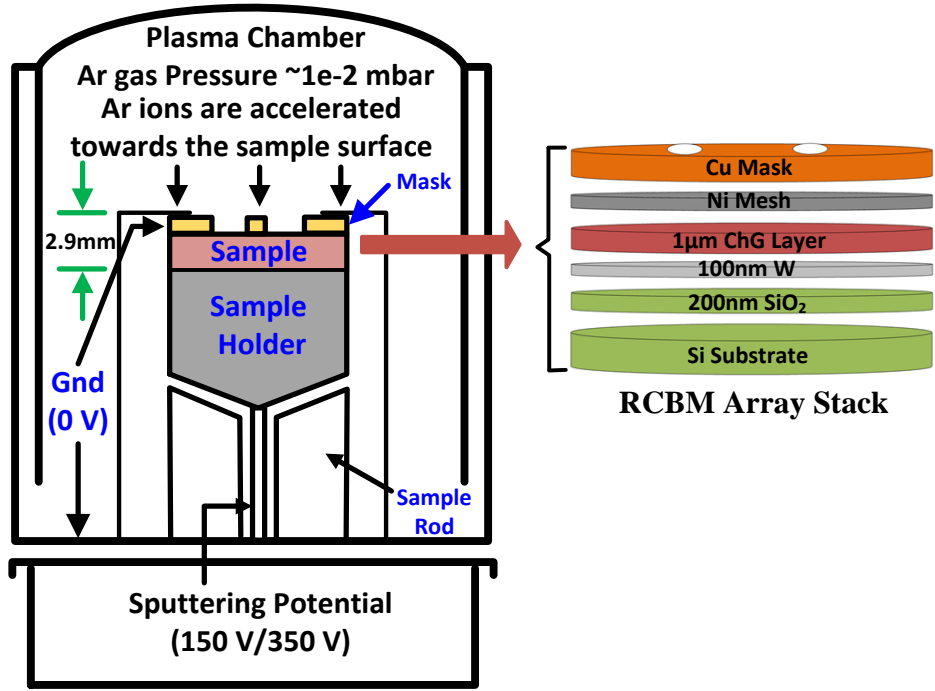

*Figure 2   SNMS System configurations with sample stack*

From the SNMS depth profile of the full sample structure the sputtering rates and bombardment times was calculated, for creating layers of 100nm and 200nm chalcogenide glass at the base of the vias that makeup the grating.  A 350 V and 150 V sputtering potential were applied to two different samples for comparison of ion-induced effects from differing bombardment energy. Each ion bombardment was performed at 100 kHz frequency with 80% duty cycle applied to the sample. Device characteristics were compared from each sputtering energy and chalcogenide layer thickness at the base of the via. 50 nm of Ag was deposited through DC magnetron sputtering with deposition pressure of $5 \times 10^{-3}$ mbar by covering one hole of copper mask after ion bombardment. The sputtering rate of Ag was calibrated by Ambios XP-1 profilometer. Analysis of the films was achieved through application of Energy Dispersive Spectroscopy (EDS) by Hitachi S-3400N EDS system. 20 kV accelerating voltage, and 10nA beam current were used for EDS measurements. The surface morphologies of Ge-Se films bombarded with $Ar^+$ ions were studied in tapping mode using OTESPA probe on a Veeco Dimensions 3100 AFM system equipped with Nanoscope IV controller. Raman spectra of the irradiated films were performed at room temperature in vacuum chamber using Horiba Jobin Yvon T64000 Raman spectroscopic system in back scattering mode.

Electrical testing on the devices were performed using an Agilent 4155B Semiconductors Parameter Analyzer equipped with triax cables to avoid residual charge build up. W and Ag pads were probed with correct biasing for SET and RESET conditions. Various cells in the array were tested in dual sweep mode with a voltage step size of 2mV and compliance current set to 50nA. Data were analyzed and recorded by Easy Expert Software provided by Agilent. The testing equipment (sample stage holder, triax cables and probes) was placed inside a Faraday cage sharing a common ground to avoid static charge build up. All electrical measurements were carried out at room temperature.

3. Results

An SEM image of the fabricated 20x20 array is shown in Figure 3 with GeSe film isolating individual cells. Energy Dispersive X-ray Spectroscopy (EDS) was performed in five different cells at five different locations on each sample, so that 25 points were used to determine the uniformity of the film within each composition. The film compositions were measured in the cell vias created by ion bombardment as well as in the planar regions shadowed by the mask and the results are presented in Table 1. Elemental mapping was also performed through EDS to determine the Ag distribution throughout the grating, as presented in Figure 4. The presence of colored regions in Si and W layer illustrates the via as the chalcogenide material is etched by ion bombardment. The increased concentration of Ge and Se, in Figure 4, corresponds to thicker chalcogenide film which is outside the via.

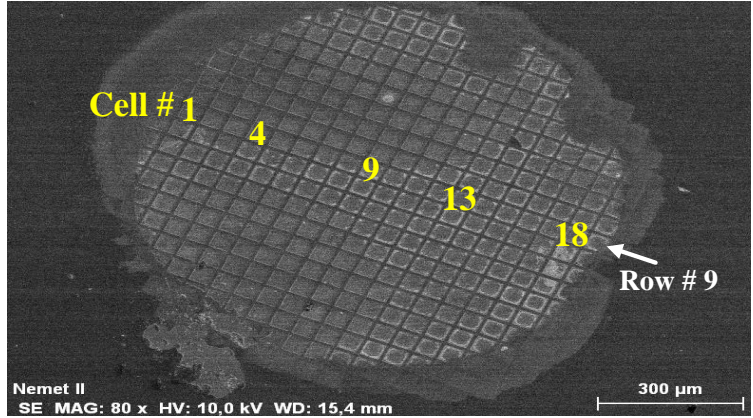

Figure 3  SEM Image of the RCBM Array

Table 1.  EDS Results on Planar and Ion Bombarded regions for $Ge_xSe_{1-x}$ Films (where x=0.2, 0.3 & 0.4)

| Sr. No. | Source Composition | Planar Region | | Ion Bombarded Region | | % Change |
|---|---|---|---|---|---|---|
| | | %Ge | %Se | %Ge | %Se | Δ |
| 1 | $Ge_{20}Se_{80}$ | 25.6±0.061 | 74.4±0.061 | 24.8 ± 0.51 | 75.2±0.51 | ± 0.8 |
| 2 | $Ge_{30}Se_{70}$ | 31.2±0.037 | 68.8±0.037 | 30.7±0.15 | 69.3±0.15 | ± 0.5 |
| 3 | $Ge_{40}Se_{60}$ | 39.9±0.19 | 60.1±0.19 | 38.8±0.14 | 61.2±0.14 | ± 1.1 |

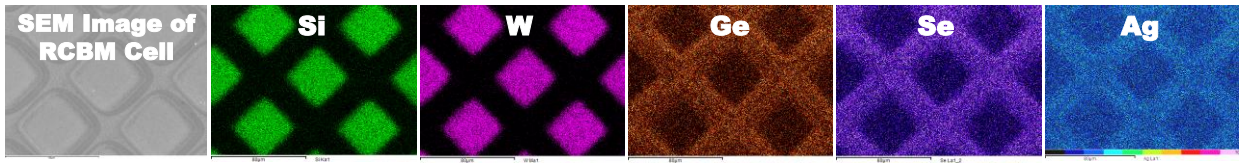

*Figure 4  Elemental mapping distribution of RCBM Array*

Raman spectra of the as deposited films, ion bombardment areas (vias), mode assignments, and corresponding structural units characteristic for $Ge_xSe_{1-x}$ (x = 0.2, 0.3 and 0.4) film compositions are presented in Figure 5a. Development of the spectra as a function of increasing Ge concentration shows an increase in the intensity of the peaks relating to the ethane-like (ETH) and the corner-shared (CS) modes when compared to the edge-shared (ES) mode. The spectra show peaks located at 178 cm$^{-1}$, 195cm$^{-1}$, and 219cm$^{-1}$ , corresponding to ETH, CS, and ES respectively [27, 28]. A close observation of ES to CS area ratio demonstrates a change in the area ratio for Ge rich compositions for the ion bombarded regions as shown in Figure 5b, with the largest change being observed in $Ge_{40}Se_{60}$, as illustrated in Figure 5c.

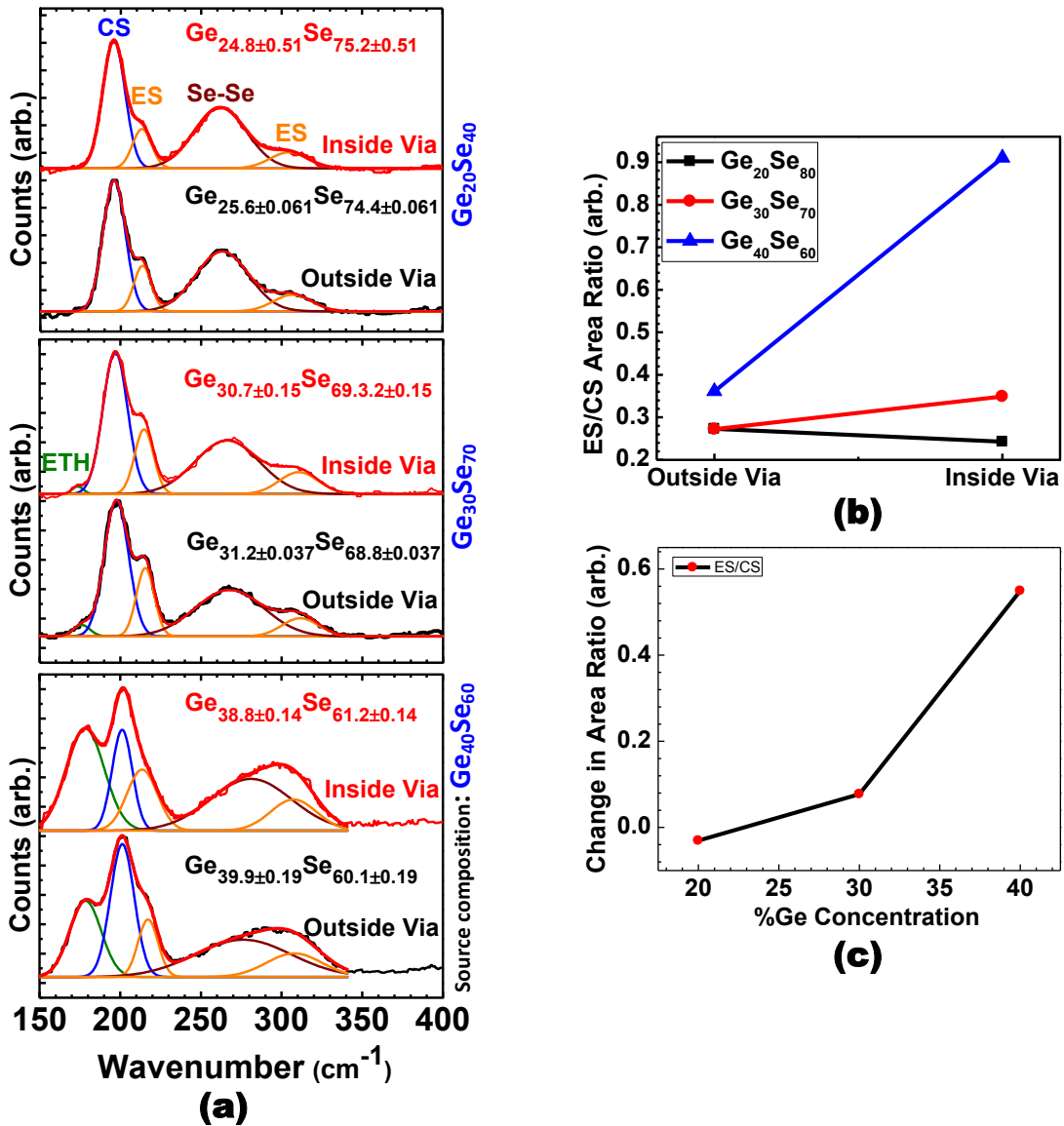

Figure 3     (a) Raman data and the corresponding mode assignment (b) Area Ratios b/w Edge shared and Corner shared modes (c) Change in area ratio with different Ge concentration

The surface morphology within the cell vias of the $Ge_xSe_{1-x}$ layer, created by Ar+ ion bombardment, were studied by AFM and the results are presented in Figure 6. AFM scans were performed on cell 1, cell 9, and cell 18 in the 9$^{th}$ row of the array structure on a 25μm$^2$ area within the device vias. An improvement in the surface smoothness can be observed in the films with increased Ge concentration.

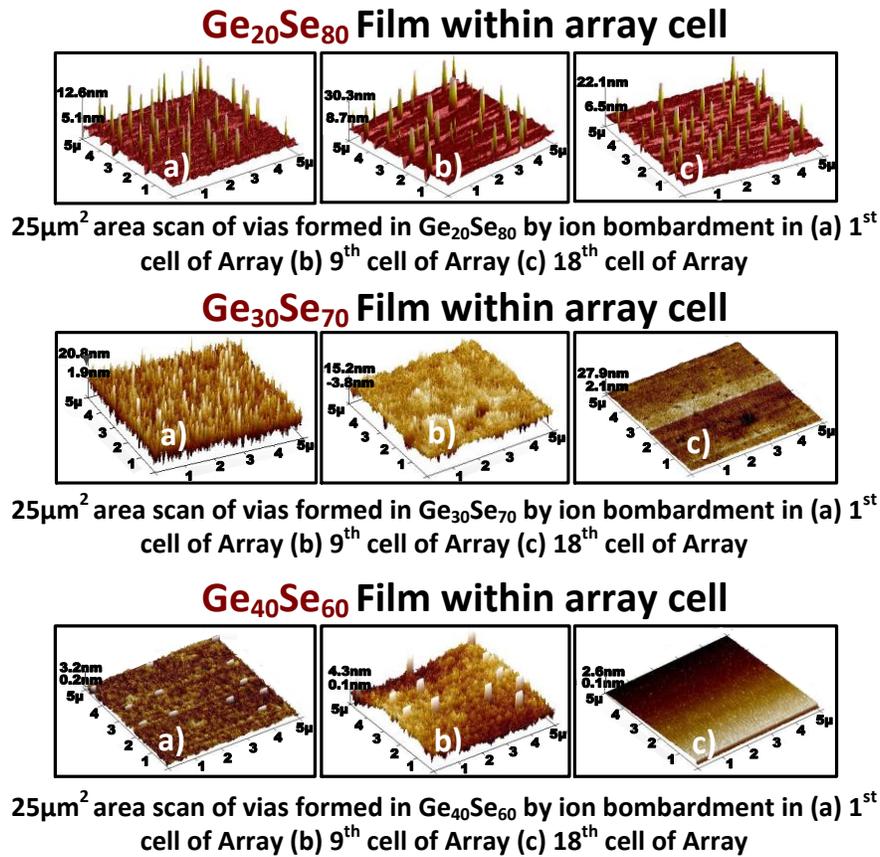

**Ge$_{20}$Se$_{80}$ Film within array cell**

25μm$^2$ area scan of vias formed in Ge$_{20}$Se$_{80}$ by ion bombardment in (a) 1$^{st}$ cell of Array (b) 9$^{th}$ cell of Array (c) 18$^{th}$ cell of Array

**Ge$_{30}$Se$_{70}$ Film within array cell**

25μm$^2$ area scan of vias formed in Ge$_{30}$Se$_{70}$ by ion bombardment in (a) 1$^{st}$ cell of Array (b) 9$^{th}$ cell of Array (c) 18$^{th}$ cell of Array

**Ge$_{40}$Se$_{60}$ Film within array cell**

25μm$^2$ area scan of vias formed in Ge$_{40}$Se$_{60}$ by ion bombardment in (a) 1$^{st}$ cell of Array (b) 9$^{th}$ cell of Array (c) 18$^{th}$ cell of Array

*Figure 4      Surface morphology of $Ge_xSe_{1-x}$ (x = 0.2, 0.3, 0.4) on a 25μm$^2$ area*

Since the performance of RCBM devices depends on formation of a conductive filament, it is important to have a smooth surface within the via for reliable device performance. EDS and AFM results suggested that amongst the studied Ge-Se compositions in this work, Ge$_{40}$Se$_{60}$ is the most suitable for RCBM array fabrication using ion bombardment method as it offers the smoothest surface. Several RCBM arrays were fabricated and tested using different mask sizes and sputtering potentials, irradiated for different time to produce different thickness at the base of the GeSe film. The current-voltage (IV) curves of different cells from fabricated arrays with different thickness of $Ge_{40}Se_{60}$ layer in the via is shown in Figure 7. Multiple IV sweeps under the same conditions were performed to ensure good endurance of the cells within the array. The devices were swept from -0.5V to 2V for each sweep. At first the current is very low (cell resistance: high) until a threshold voltage of ~0.9 V is exceeded. At that moment a conductive connection is formed between the top and bottom electrodes causing a steep increase in the current until it reaches the compliance current, limited to 50nA (cell resistance: low). The compliance current was set to protect the device during switching. Analysis of these results shows that the best device performance is achieved with a sputtering potential of 150V and 100nm thick $Ge_{40}Se_{60}$ layer left in the device active area after sputtering.

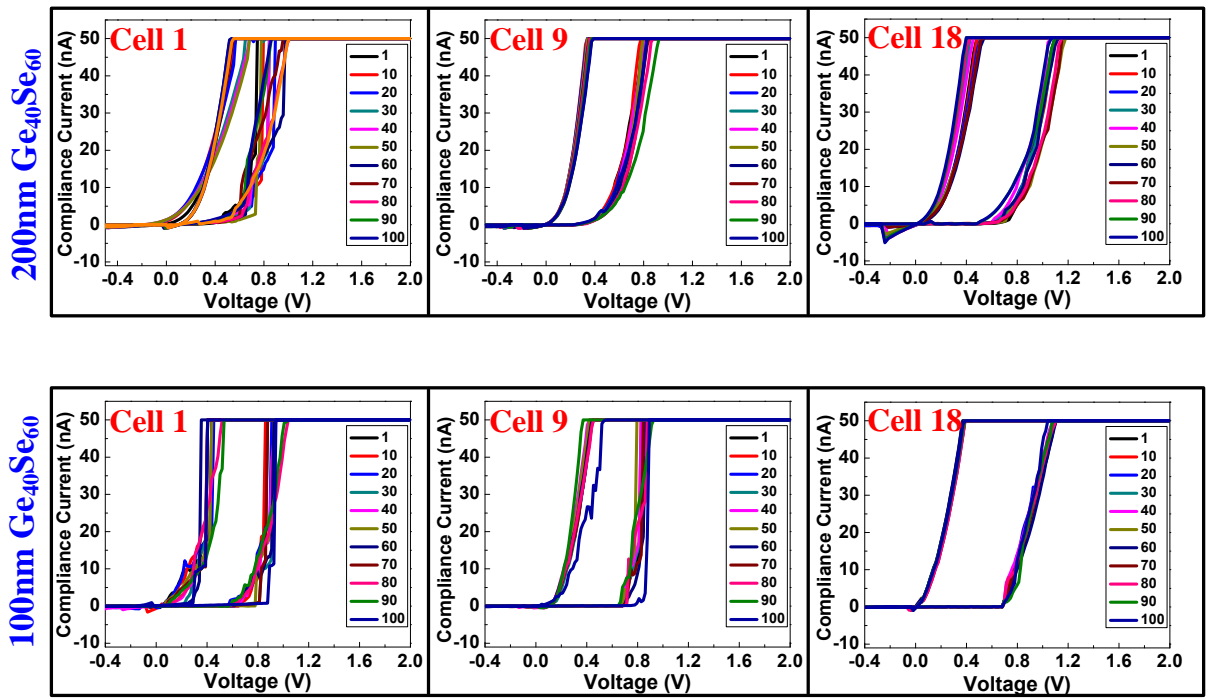

*Figure 5        IV curves in different cells of the 9th row for the fabricated RCBM array*

## 4. Discussion:

Compared to electrons, accelerated ions displace larger atoms in the target material due to their larger mass. The accelerated $Ar^+$ ions used to form the vias in the $Ge_xSe_{1-x}$ film alter the film composition, which further progresses in structural changes and eventually affects the device performance. The two processes of compositional and structural changes taking place by ion bombardment, are imperative for understanding thin films stability and hence device performance. The former is elucidated through EDS analysis of the bare and ion bombarded region, and the latter is investigated by studying Raman Spectroscopy.

Analysis of the Raman spectra using the area ratio between the Edge shared (ES) structural units corresponding to the peak located at 218 cm$^{-1}$ and Corner shared (CS) structural units corresponding to the peak at 202 cm$^{-1}$, demonstrates an increased destruction of ES structures and subsequent increase in the CS structures in Ge rich compositions ($Ge_{30}Se_{70}$/$Ge_{40}Se_{60}$) and vice versa in Se rich compositions ($Ge_{20}Se_{80}$). This suggests that the $Ar^+$ ions mainly affect the bonding sites between Ge-Se atoms. Due to the decreased availability of CS structural units in Ge-rich films, this bond is likely associated with an ES unit. Annihilation of the Ge-Se bond causes the destruction of the ES unit, creating a negative charge on the Se atom. Thus, the previously detectable ES structural units are removed from the system and their intensity in the Raman Spectra decreases. The maximum change in the area ratios is observed in $Ge_{40}Se_{60}$ composition where this effect is most dominating.

The performance of RCBM devices is based on the growth of the filament, and a smooth surface is very important for providing sites for nucleation and growth of the conductive molecular bridge. A tendency in the surface roughness improvement is observed with increasing Ge concentration, mainly due to the packing fraction [29] associated with these glass networks. A close observation on the height bar of AFM scans illustrates huge hillocks in $Ge_{20}Se_{80}$ composition which is associated to loosing of Se bonds. Lowest loss of Se occurs in $Ge_{30}Se_{70}$ with largest surface roughness mainly due to decrease in homopolar bonds. Ion bombardment on $Ge_{40}Se_{60}$ film results in decrease of CS bonds due to an extraction of Se atoms from the corner shared Se atoms, resulting in formation of Se-Se bond. Lowest surface roughness is observed in $Ge_{40}Se_{60}$ composition since the major backbone of the structure remains undamaged. Because of this, the devices were fabricated and their performance was electrically tested with $Ge_{40}Se_{60}$ composition.

An important consideration with this lithography-free method for array formation is how the device performance alters with the thickness of the ChG layer at the base of the via. The device performance is evaluated with 100nm and 200nm thick $Ge_{40}Se_{60}$ film in the via. The devices with 100nm $Ge_{40}Se_{60}$ film at the base show lower threshold voltage (~0.8) with good uniformity and repeatability. The IV curves in Figure 7 suggest that the array devices with 100nm $Ge_{40}Se_{60}$ films turn on faster (instantaneously) compared to devices with 200nm $Ge_{40}Se_{60}$ in the via. The two resistive states i.e. LRS and HRS are presented in Figure 8 which shows 3 orders of magnitude difference between the ON and OFF state. A slightly lower ON state resistance is observed within the array devices with 100nm of $Ge_{40}Se_{60}$ film due to comparatively faster switching and lower threshold voltage.

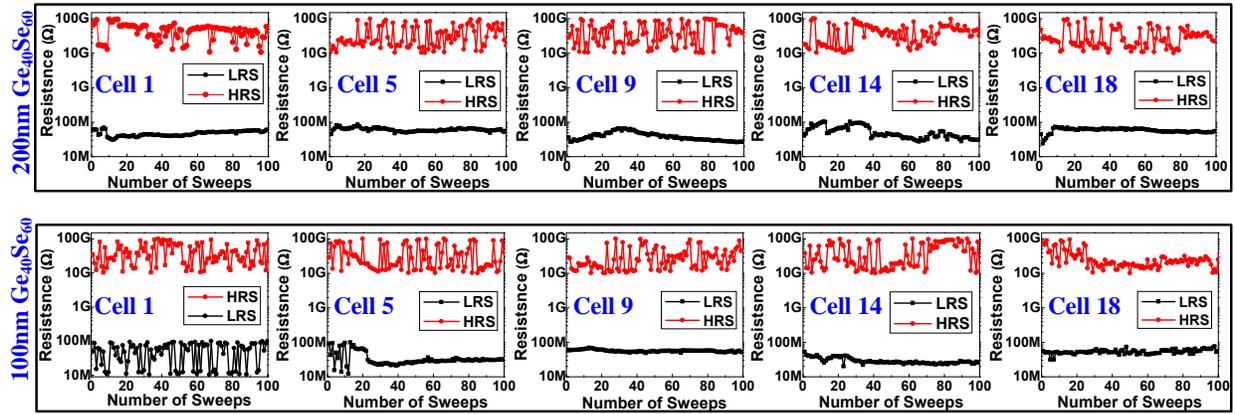

*Figure 8.    HRS and LRS plot of Ge$_{40}$Se$_{60}$ film with different thickness in the via*

## 5. Conclusion

In this work, we studied the effect of the Ar$^+$ ion on the GeSe system and successfully demonstrated the fabrication of a lithography free RCBM array with individual cell addressing. In addition, data has been presented about the electrical performance of the cells within RCBM array, based on different thickness of GeSe layer in the active region. It was found that the Ge rich film offered the least surface roughness which contributed to the device performance, and is attributed to the formation of rigid structure and the availability of Ge-Ge bonds. Further improvement in the cell performance can be achieved by filling the via homogenously with Ag.

**Acknowledgement**

This work has been supported by IMI-NFG, Lehigh University through National Science Foundation Grant DMR 0844014 and by the European Union and the State of Hungary under Grant # TÁMOP 4.2.4. A/2-11-1-2012-0001 (National Excellence Program) and TÁMOP-4.2.2.A-11/1/KONV-2012-0036, which are co-financed by the European Union and the European Social Fund. The funding of Defense Threat Reduction Agency (DTRA) under grant no: HDTRA1-11-1-0055 is also acknowledged. The authors would also like to thank Dr. James Reed of DTRA for his support. The authors would also like to acknowledge the Surface Science

Lab at Boise State University for AFM use and Dr. Paul Davis for assistance in performing AFM.

**References:**

[R. Lovics] R. Lovics, A. Csik, V. Takáts, J. Hakl, K. Vad, G.A. Langer, "Depth profile analysis of solar cells by Secondary Neutral Mass Spectrometry using conducting mesh" Vacuum 86 (2012) 721-723

[1]     M. T. Bohr, "Nanotechnology goals and challenges for electronic applications," *Nanotechnology, IEEE Transactions on,* vol. 1, pp. 56-62, 2002.
[2]     Y. Koh, "NAND Flash scaling beyond 20nm," in *Memory Workshop, 2009. IMW'09. IEEE International*, 2009, pp. 1-3.
[3]     G. E. Moore, "Progress in digital integrated electronics," in *Electron Devices Meeting, 1975 International*, 1975, pp. 11-13.
[4]     G. E. Moore, "Cramming more components onto integrated circuits, Reprinted from Electronics, volume 38, number 8, April 19, 1965, pp.114 ff," *Solid-State Circuits Newsletter, IEEE,* vol. 11, pp. 33-35, 2006.
[5]     K. Aratani, K. Ohba, T. Mizuguchi, S. Yasuda, T. Shiimoto, T. Tsushima*, et al.*, "A novel resistance memory with high scalability and nanosecond switching," in *Electron Devices Meeting, 2007. IEDM 2007. IEEE International*, 2007, pp. 783-786.
[6]     R. Waser, R. Dittmann, G. Staikov, and K. Szot, "Redox-Based Resistive Switching Memories - Nanoionic Mechanisms, Prospects, and Challenges," *Advanced Materials,* vol. 21, pp. 2632-+, Jul 2009.
[7]     A. Sawa, "Resistive switching in transition metal oxides," *Materials today,* vol. 11, pp. 28-36, 2008.
[8]     R. Waser and M. Aono, "Nanoionics-based resistive switching memories," *Nature materials,* vol. 6, pp. 833-840, 2007.
[9]     H. Akinaga and H. Shima, "Resistive random access memory (ReRAM) based on metal oxides," *Proceedings of the IEEE,* vol. 98, pp. 2237-2251, 2010.
[10]    M. N. Kozicki, M. Park, and M. Mitkova, "Nanoscale memory elements based on solid-state electrolytes," *Nanotechnology, IEEE Transactions on,* vol. 4, pp. 331-338, 2005.
[11]    S. Seo, M. J. Lee, D. H. Seo, E. J. Jeoung, D. S. Suh, Y. S. Joung*, et al.*, "Reproducible resistance switching in polycrystalline NiO films," *Applied Physics Letters,* vol. 85, pp. 5655-5657, 2004.
[12]    J. F. Gibbons and W. E. Beadle, "Switching properties of thin NiO films," *Solid-State Electronics,* vol. 7, pp. 785-790, 11/ 1964.
[13]    O. Kavehei, C. Kyoungrok, L. Sangjin, K. Sung-Jin, S. Al-Sarawi, D. Abbott*, et al.*, "Fabrication and modeling of Ag/TiO$_2$/ITO memristor," in *Circuits and Systems (MWSCAS), 2011 IEEE 54th International Midwest Symposium on*, 2011, pp. 1-4.
[14]    B. J. Choi, D. S. Jeong, S. K. Kim, C. Rohde, S. Choi, J. H. Oh*, et al.*, "Resistive switching mechanism of TiO$_2$ thin films grown by atomic-layer deposition," *Journal of Applied Physics,* vol. 98, pp. 033715-033715-10, 2005.
[15]    Z. Fang, H. Yu, W. Liu, Z. Wang, X. Tran, B. Gao*, et al.*, "Temperature Instability of Resistive Switching on {HfO} _ {x}-Based RRAM Devices," *IEEE Electron Device Letters,* vol. 31, pp. 476-478, 2010.
[16]    H. Lee, P. Chen, T. Wu, Y. Chen, C. Wang, P. Tzeng*, et al.*, "Low power and high speed bipolar switching with a thin reactive Ti buffer layer in robust HfO2 based RRAM," in *Electron Devices Meeting, 2008. IEDM 2008. IEEE International*, 2008, pp. 1-4.
[17]    Y.-S. Chen, T.-Y. Wu, P.-J. Tzeng, P.-S. Chen, H.-Y. Lee, C.-H. Lin*, et al.*, "Forming-free HfO$_2$ bipolar RRAM device with improved endurance and high speed operation," in *VLSI Technology, Systems, and Applications, 2009. VLSI-TSA'09. International Symposium on*, 2009, pp. 37-38.

[18] M. Mitkova and Y. Sakaguchi, "Nano-ionic nonvolatile memory devices–basic ideas and structural model of rigid Ge-S glasses as medium for them," 2009.
[19] M. Mitkova, Y. Sakaguchi, D. Tenne, S. K. Bhagat, and T. L. Alford, "Structural details of Ge-rich and silver-doped chalcogenide glasses for nanoionic nonvolatile memory," *physica status solidi (a),* vol. 207, pp. 621-626, 2010.
[20] K.-H. Kim, S. Gaba, D. Wheeler, J. M. Cruz-Albrecht, T. Hussain, N. Srinivasa*, et al.*, "A functional hybrid memristor crossbar-array/CMOS system for data storage and neuromorphic applications," *Nano letters,* vol. 12, pp. 389-395, 2011.
[21] X. Hu, S. Duan, L. Wang, and X. Liao, "Memristive crossbar array with applications in image processing," *Science China Information Sciences,* vol. 55, pp. 461-472, 2012.
[22] S. Yu, B. Gao, Z. Fang, H. Yu, J. Kang, and H.-S. P. Wong, "A neuromorphic visual system using RRAM synaptic devices with Sub-pJ energy and tolerance to variability: Experimental characterization and large-scale modeling," in *Electron Devices Meeting (IEDM), 2012 IEEE International*, 2012, pp. 10.4. 1-10.4. 4.
[23] K. Bickerstaff and E. Swartzlander, "Memristor-based arithmetic," in *Conference Record of the Forty Fourth Asilomar Conference on Signals, Systems and Computers (ASILOMAR)*, 2010, pp. 1173-1177.
[24] B. Mouttet, "Proposal for memristor crossbar design and applications," in *Memristors and Memristive Systems Symposium, UC Berkeley*, 2008.
[25] J. Cong and B. Xiao, "mrFPGA: A novel FPGA architecture with memristor-based reconfiguration," in *IEEE/ACM International Symposium on Nanoscale Architectures (NANOARCH)*, pp. 1-8, 2011.
[26] R. Kundu, K. Bhatia, N. Kishore, P. Singh, and C. Vijayaraghavan, "Effect of addition of Zn impurities on the electronic conduction in semiconducting $Se_{80-x}Te_{20}Zn_x$ glasses," *Philosophical Magazine B,* vol. 72, pp. 513-528, 1995.
[27] K. Jackson, A. Briley, S. Grossman, D. V. Porezag, and M. R. Pederson, "Raman-active modes of a-$GeSe_2$ and a-$GeS_2$: A first-principles study," *Physical Review B,* vol. 60, p. R14985, 1999.
[28] M. T. Shatnawi, C. L. Farrow, P. Chen, P. Boolchand, A. Sartbaeva, M. Thorpe*, et al.*, "Search for a structural response to the intermediate phase in Ge_ {x} Se_ {1− x} glasses," *Physical Review B,* vol. 77, p. 094134, 2008.
[29] P. Chen, M. Mitkova, D. A. Tenne, K. Wolf, V. Georgieva, and L. Vergov, "Study of the Sorption Properties of $Ge_{20}Se_{80}$ thin Films for $NO_2$ Gas Sensing," *Thin Solid Films,* 2012.